\documentclass[preprint,aps]{revtex4}

\usepackage{graphicx}
\usepackage{dcolumn}
\usepackage{bm}

\begin{document}
\title{Achieving Perfect Imaging beyond Passive and Active Obstacles by a Transformed Bilayer Lens}
\date{\today}
\author{Wei Yan, Min Yan}
\author{Min Qiu}
\email{min@kth.se} \affiliation{Laboratory of Optics, Photonics and
Quantum Electronics, Department of Microelectronics and Applied
Physics, Royal Institute of Technology (KTH), Electrum 229, 16440
Kista, Sweden}
\date{\today}
\begin{abstract}
A bilayer lens is proposed based on transformation optics. It is
shown that Pendry's perfect lens, perfect bilayer lens made of
indefinite media, and the concept of compensated media are well
unified under the scope of the proposed bilayer lens. Using this
concept, we also demonstrate how one is able to achieve perfect
imaging beyond passive objects or active sources which are present
in front of the lens.

\end{abstract}
\maketitle The form-invariance property of the Maxwell's equations
in any coordinate system \cite{P1,Ulf1,Ulf2} provides us a
convenient guideline to design the so-called transformation media
for controlling electromagnetic (EM) fields or light in an
unprecedented manner. In this design methodology, a coordinate
transformation function describes the desired trajectory of EM field
in a direct geometrical meaning \cite{Ulf1}, and it determines the
material parameters of the transformation media by tensor rules
\cite{P1,Ulf1}. The most intriguing application of the theory is
invisibility cloak, as proposed by Pendry et al \cite{P1} and
Leonhardt et al \cite{Ulf2} in their preliminary works. Since then,
this particular field of study, referred to as transformation
optics, has received intense attention from the optics community. We
have so far observed a surge of theoretical discussions
\cite{Cai,Cum,Ruan,Chen1,Ulf3,Yan} and even experimental efforts
\cite{Sch} related especially to the cloaking subject. Apart from
invisibility cloaks, some other novel applications of transformation
optics, such as, beam shifters and splitters \cite{Rahm}, field
rotation \cite{Chen2}, and electromagnetic wormholes \cite{Gf}, have
been proposed.

In Refs. [9], it is shown that a Pendry's perfect slab lens
\cite{P2} with $n=-1$ can be interpreted by transformation optics.
In particular, the lens body together with its free space background
can be considered as transformation media obtained from a coordinate
transformation of free space based on a folded mapping function.
Following this strategy, a perfect cylindrical (spherical) shape
lens can be designed by deploying a folded radial spatial mapping in
a cylindrical (spherical) coordinate system \cite{YM2}. On the other
hand, Pendry's perfect slab lens can be understood as a special
example of compensated media \cite{P3}. In this paper, by studying a
transformed bilayer structure, we notice that transformation optics
provides a clear physical interpretation of such compensated media,
the compensated media are just special examples of the transformed,
in particular, bilayer structure. Thus, we can define a generalized
concept of compensated media, which well unify Pendy's slab lens and
the indefinite media lens proposed by Smith et al \cite{Smith,
Sch2}. Based on this understanding, we show that perfect imaging can
be realized even when passive obstacles or active emitters are
obstructing our object to be imaged.

Consider a coordinate transformation which transforms a single slab
structure placed in EM space into a bilayer slab structure, as
illustrated in Fig. 1. The single slab structure in EM space has
relative permittivity $\epsilon$ and relative permeability $\mu$,
and may contain active sources denoted by $\overline J$ and $\rho$.
The transformation function is also illustrated in Fig. 1. It is
seen that the region $ z^{'}\in [d,e]$ in EM virtual space
transforms into two regions $ z\in [a,b]$ and $ z\in [b,c]$ in
physical space following transformation functions $z^{'}=f(z)$ and
$z^{'}=g(z)$, respectively. The two transformation functions have
the boundary conditions $f(a)=g(c)=d$ and $f(b)=g(b)=e$. The
material parameters as well as sources in the transformed layers,
denoted by layer I and layer II, can be expressed as \cite{P1,Ulf1}
\begin{equation}
\varepsilon _1  = \det (\Lambda _f )^{-1}\Lambda _f  \varepsilon
\Lambda _f ,\;\mu _1  = \det (\Lambda _f )^{-1}\Lambda _f  \mu
\Lambda _f,
\end{equation}
\begin{equation}
\varepsilon _2  = \det (\Lambda _g )^{-1}\Lambda _g \varepsilon
\Lambda _g ,\;\mu _2  = \det (\Lambda _g)^{-1}\Lambda _g  \mu
\Lambda _g,
\end{equation}
\begin{equation}
\overline J_1=\det (\Lambda _f )^{-1}\Lambda _f J ,\; \rho_1=\det
(\Lambda _f )^{-1}\rho,
\end{equation}
\begin{equation}
\overline J_2=\det (\Lambda _g )^{-1}\Lambda _g J ,\; \rho_2=\det
(\Lambda _g )^{-1}\rho,
\end{equation}
where $"\det"$ represents the determinant of a matrix, $\Lambda _f =
diag[1,1,1/f^{'}(z)]$ and $ \Lambda _g  = diag[1,1,1/g^{'}(z)]$.

If the fields in the layer I are denoted by $\overline E_1(x,z)$ and
$\overline H_1(x,z)$, based on transformation optics, the fields in
the layer II should be related to the fields in the layer I by
\begin{equation}
\overline E _2 (x,y,z) = diag[1,1,g^{'} (z)/f^{'} (z_m )]\overline E
_1 (x,y,z_m ),
\end{equation}
\begin{equation}
\overline H _2 (x,y,z) = diag [1,1,g^{'} (z)/f^{'} (z_m )]\overline
H_1 (x,y,z_m ),
\end{equation}
where $z\in [b,c]$ and $z_m\in [a,b]$ correspond to the same $z^{'}$
. At boundaries $z=a$ and $z=c$, we have $\overline E _2 (x,c) =
diag[1,1,h^{'} (c)/g^{'} (a )]\overline E _1 (x,a)$ and $\overline H
_2 (x,c) = diag[1,1,h^{'} (c)/g^{'} (a )]\overline H _1 (x,a)$.
Therefore the tangential fields at $z=a$ and $z=c$ two boundaries
have the same values. The left and right outside backgrounds appear
as connecting each other directly. Here, we note that this
phenomenon is independent of the background material choice, or
whether active sources are involved in transformation. Effectively,
the two boundaries $z=a$ and $z=c$ of the bilayer are perfect
duplicates of the $z^{'}=d$ boundary of the original slab. Thus, if
the bilayer is put into a homogenous background, the fields will
perfectly tunnel through the bilayer without any reflection.

Now we examine the material properties of such a bilayer lens in
general. Consider that transformation functions are linear functions
with slopes being $p_1$ and $-p_2$, respectively, where $p_1p_2>0$.
Thus, the thicknesses of layer I and layer II (denoted by $L_1$ and
$L_2$ respectively) relate to each other by $L_1=\gamma L_2$, where
$\gamma =p_2/p_1$. Observing Eqs. (5) and (6), we have $\varepsilon
_2 (x,y,z) = - \gamma diag[1,1, - 1/\gamma ]\varepsilon _1(x,y,z_m
)diag[1,1, - 1/\gamma ]$, $\mu _2 (x,y,z) = - \gamma diag[1,1, -
1/\gamma ]\mu _1(x,y,z_m )diag[1,1, - 1/\gamma ]$, where $z$ and
$z_m$ are defined as the same as in Eqs. (5) and (6). When
$\gamma=1$, the above equations describe exactly the complementary
media proposed by Pendry in Ref. [16]. In particular, $\varepsilon
_2 (x,y,z)=-\varepsilon _1 (x,y,z_m)$ and $\mu _2 (x,y,z)=-\mu _1
(x,y,z_m)$, if $\varepsilon _{(1,2)}$ and $\mu _{(1,2)}$ are
diagonal matrices. As discussed in Ref. [16], the bilayer composed
by two complimentary layers with the same thickness can transfer EM
fields perfectly from one interface to the other. Here, we show that
transformation optics facilitates simple and clear geometrical
interpretation of the perfect lensing phenomenon of such
complementary media. In fact, bilayer structure as complementary
media can be extended to the following situations: (1) $\gamma\ne
1$, and/or (2) $f(z)$ and $g(z)$ are nonlinear functions, and/or (3)
active sources are embedded in the bilayer.

As special examples, here we show how the transformed bilayer lens
can encompass Pendry's perfect lens as well as the bilayer
indefinite media lens proposed by Smith et al. Consider the single
slab in EM space is homogenous. With linear coordinate
transformations, the individual layers of the corresponding bilayer
in physical space are therefore also homogenous. If the parameters
of layer I are denoted by
$\epsilon_1=diag[\epsilon_{1x},\epsilon_{1y},\epsilon_{1z}]$ and
$\mu_1=diag[\mu_{1x},\mu_{1y},\mu_{1z}]$, the parameters of the
layer II are
$\epsilon_2=-diag[\epsilon_{1x}\gamma,\epsilon_{1y}\gamma,\epsilon_{1z}/\gamma]$,
$\mu_2=-diag[\mu_{1x}\gamma,\mu_{1y}\gamma,\mu_{1z}/\gamma]$.
Consider Pendry's perfect lens with $\epsilon_1=\mu_1=-1$ \cite{P2},
which is located in $z\in[0,S]$ and is put in free space background.
The imaging process of this lens can be understood easily if we
interpret the system as two connected bilayer structures for the
billayer 1 with $\epsilon_1=-\epsilon_2=1$, $\mu_1=-\mu_2=1$ and
$\gamma=1$, and the bilayer 2 with $-\epsilon_1=\epsilon_2=1$,
$-\mu_1=\mu_2=1$ and $\gamma=1$, as illustrated in Fig. 2. Refer to
the figure, if a source is put on the plane $z=-W$ with $W<S$, its
perfect image (Image 1) will be formed at $z=W$ through perfect
field tunneling by the bilayer 1. Then, through field tunneling by
the bilayer 2, a perfect image 2 is constructed in free space at
$z=2S-W$. Furthermore, consider the single slab in EM space is made
of indefinite medium. Subject to linear coordinate transformations,
the corresponding bilayer is composed by homogenous indefinite media
where the material tensors $\epsilon_{(1,2)}$ and $\mu_{(1,2)}$ have
both positive and negative components. In this case, we notice that
such an indefinite bilayer is exactly the perfect lens made of the
indefinite bilayer proposed by Smith et al. \cite{Smith,Sch2}.

Now we give a further example of a homogenous bilayer lens located
at $z\in[0, 4\lambda]$ with $\epsilon_1=4$, $\mu_1=1$,
$\epsilon_2=-4+0.008i$ and $\mu_2=-1+0.008i$, where $\lambda$ is the
operating wavelength. The simulations are carried out with the
finite-element method (FEM) using the commercial COMSOL Multiphysics
package. Notice that in simulations presented in this paper, all
material parameters smaller than $1$ are all given a small imaginary
part of $0.008i$. This small imaginary part not only avoids the
theoretical singularity problem, but also physically represents
losses possessed by realistic metamaterails. The electric field
distribution for a line current source
$J_s=A\delta(z+0.01\lambda)\delta(x)$ interacting with the bilayer
is plotted in Fig. 3 (a). It is seen that the EM field tunnels
through the bilayer perfectly, and  an almost perfect image is
achieved at the exit boundary $z=4\lambda$. In Fig. 3(b), the field
intensity distribution as a function of lateral position at the exit
boundary $z=4\lambda$, for both with and without the bilayer, are
plotted for comparison. The image constructed by the bilayer
achieves a subwavelength resolution with a full width at half
maximum (FWHM) of $0.28\lambda$. Whereas for the case without the
bilayer, the field has decayed significantly and appears
featureless.

If the single slab in EM space is inhomogeneous, the corresponding
transformed bilayer will be inhomogeneous too. This hints that
perfect image can be achieved even when obstacles are obstructing
the object to be imaged. To illustrate this idea, we consider that a
dielectric cylindrical obstacle, with a radius of $2/3\lambda$ and a
refractive index of $2$, is put at at ($x=0$ $z=\lambda$), just in
front of a Pendy's lens. The lens has $\epsilon_L=-1+0.008i$,
$\mu_L=-1+0.008i$, and is positioned at $z\in[0,10/3\lambda]$. The
simulated electric field distribution for a line current source
$J_s=A\delta(z+7/3\lambda)\delta(x)$ interacting with the lens and
obstacle is plotted in Fig. 4(b). In Fig. 4(a), we also plot the
electric field distribution when the cylindrical obstacle is absent.
The Pendry's lens is outlined by solid lines, while dashed lines
outline two bilayers. Comparing Figs. 4(a) and (b), it is clearly
seen that the images are distorted by the obstacle, since the up
bilayer fails to transfer the field perfectly. To overcome this
problem, we embed an complementary cylinder inside the lens, which
has a permittivity of $-4+0.008i$ and a permeability of $-1+0.008i$.
The complementary cylinder is positioned.symmetrically with the
outside obstacle about $z=0$. Thus, the up bilayer works as a
self-compensating bilayer lens again. In Fig. 4(c), we show the
electric field distribution when the complementary cylinder is added
inside the lens. Two images are constructed almost perfectly again.
In Fig. 4(d), the field intensity distributions at the image plane
$z=13/3\lambda$ for Figs. 4(a), (b) and (c) are plotted. It is
observed that the intensity curves for (a) and (c) agree with each
other quite well.

Next, we analysis the situation where active sources presented in
the single layer in EM space. The transformed bilayer thus contains
double amount of the mapped sources. The currents or changes of the
mapped sources are determined through Eqs. (3-4). Here we assume
that no other sources are present in the background. EM fields are
therefore radiated only by the sources embedded in the bilayer.
Thus, the power flows cross two boundaries $z=a$ and $z=c$ should be
in opposite directions, i.e., outward from the bilayer. However, as
discussed previously, tangential EM fields at $z=a$ and $z=c$
boundaries are always the same, which indicates that the power flows
cross $z=a$ and $z=c$ should be of the same value and with the same
direction. The above two conclusions are contradictory unless we
acknowledge that the power flows cross $z=a$ and $z=c$ are both
zero. It follows that no power flow propagates cross both $z=a$ and
$z=c$. Therefore, EM fields outside the bilayer should be zero or
they consist only evanescent components. However, if the fields
outside the bilayer indeed consist evanescent components, they must
decay in the same direction on both sides of the bilayer, since the
EM tangential fields at two boundaries $z=a$ and $z=c$ have the
identical values. This leads to infinite EM evanescent fields at
$z=-\infty$ or $z=+\infty$, which is obviously unphysical. Hence,
one comes to the conclusion that the EM fields outside the bilayer
are completely zero. An outside observer can't see any source
embedded in the bilayer. As an example to illustrate this statement,
we consider that a bilayer is located at $z\in[-2\lambda,2\lambda]$
which has the same material parameters as in Fig. 3(a). Two line
current sources $\overline J_1=A\delta(x)\delta(z+1.9\lambda)\hat y$
and $\overline J_2=-A\delta(x)\delta(z-1.9\lambda)\hat y$ are
embedded in layer I and II, respectively. The simulated electric
field distribution is plotted in Fig. 5 (a). It is clearly seen that
EM fields are nearly zero outside the bilayer. For a comparison, we
also plot the electric field distribution when $J_2=0$ in Fig. 5(b),
where relatively large fields outside the bilayer are observed.

Such a bilayer with active sources can be applied to design a
perfect lens for achieving subwavelength imaging beyond active noise
sources. To illustrate this idea, we consider that a sheet current
$J_s=4A\delta(x+z/6\lambda)\;x\in[-2\lambda,2\lambda]$ is put in the
front of the lens described in Fig. 4(a). The corresponding electric
field distribution is plotted in Fig. 6(a). It is seen that the
original clear images can't be distinguished anymore due to the
sheet source. To overcome this problem, we embed a complementary
sheet source within the lens to cancel out the outside noise source.
The complementary source and the outside noise are symmetrical about
$z=0$, while their phases have a $\pi$ difference. The electric
field distribution for this case is plotted in Fig. 6(b). Clear
images reemerge. In Fig. 6(c), we plot the field intensity
distributions along the image plane $z=13/3\lambda$ for both Fig.
6(a) and (b). Also, the field intensity at the image plane for Fig.
4(a) is imposed as a reference. Clearly, the intensity curves for
Fig. 6(b) and Fig. 4(a) agree with each other almost perfectly.

In conclusion, we studied bilayer slab structures obtained by
coordinate-transforming a single slab in EM space. The tangential
fields at the two boundaries of such a bilayer structure have the
same values, which is independent of the background material or the
existence of active sources involved in the transformation. The
transformed active sources in the bilayer radiate no EM fields into
the outside background. If the bilayer structure is put in a
homogenous background, it will operate as a perfect tunneling lens,
which relays EM fields from one boundary to the other perfectly.
Considering the material parameters of the bilayer structure, we
find the perfect NIM lens, perfect indefinite media lens, and
complementary media are well unified under the system of
transformation optics. Based on this understanding, constructions of
perfect bilayer lenses beyond passive and active obstacles are
demonstrated. We envision that the idea of bilayer slab lens can
also be extended to design bilayer lenses in other geometries, such
as cylindrical and spherical ones.
\section*{Acknowledgements} This work is supported by the Swedish Foundation for Strategic Research (SSF)
through the Future Research Leaders program, the SSF Strategic
Research Center in Photon- ics, and the Swedish Research Council
(VR).

\newpage
\section*{Figure captions}
\textbf{Figure 1}: Illustration of a bilayer slab structure obtained
by coordinate transformations from a single slab layer.

\textbf{Figure 2}: Illustration of Pendry's perfect lens from the
view of the bilayer structure.

\textbf{Figure 3}: (a) Electric field distribution for a line
current source $J_s=A\delta(z+0.01\lambda)\delta(x)$ interacting
with a bilayer lens located in $z\in[0, 4\lambda]$; (b) Field
intensities at the exit boundary $z=4\lambda$ for the cases with the
bilayer and without bilayer. The bilayer lens has $\epsilon_1=4$,
$\mu_1=1$, $\epsilon_2=-4+0.008i$ and $\mu_2=-1+0.008i$ and
$L_1=L_2=2\lambda$.

\textbf{Figure 4}: (a) Electric field distribution for a line
current source $J_s=A\delta(z+7/3\lambda)\delta(x)$ interacting with
a Pendry's lens $\epsilon_L=-1+0.008i$ and $\mu_l=-1+0.008i$; (b)
electric field distribution when a dielectric cylinder is put in the
front of the lens; (c) electric field distribution when the
dielectric cylinder together with its compensated cylinder are put
outside and inside the lens, respectively. (d) Field intensity at
the image plane $z=13/3\lambda$ for (a), (b) and (c). The dielectric
cylinder has a radius of $2/3\lambda$ with the center at
$z=-\lambda$ and $x=0$, and a refractive index of $2$.

\textbf{Figure 5}: Electric field distribution for (a) two current
sources $\overline J_1=A\delta(x)\delta(z+1.9\lambda)\hat y$ and
$\overline J_2=-A\delta(x)\delta(z-1.9\lambda)\hat y$ embedded in
the bilayer ; (b) only $\overline J_1$ is embedded in the bilayer.
The bilayer lens has the same material parameters as in Fig. 3.

\textbf{Figure 6}: Electric field distribution when (a) a current
sheet $J_s=4A\delta(x+z/6\lambda)\;x\in[-2\lambda,2\lambda]$ is put
in the front of the lens of Fig. 4(a); (b) the compensated current
sheet is put inside the lens. (c) Field intensity at the the image
plane $z=13/3\lambda$ for (a), (b), and also Fig. 4(a).

\newpage
\begin{figure}[htbp] \centering
\includegraphics[width=12cm]{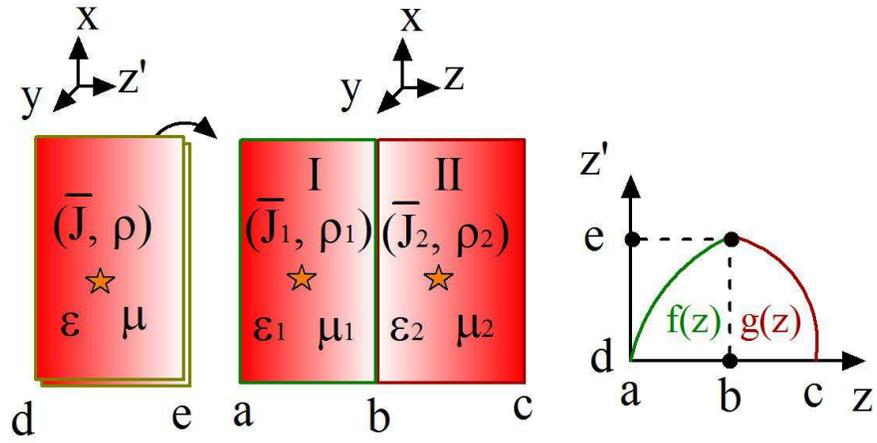}
\caption{Illustration of a bilayer slab structure obtained by
coordinate transformations from a single slab layer.}
\end{figure}
\newpage
\begin{figure}[htbp] \centering
\includegraphics[width=10cm]{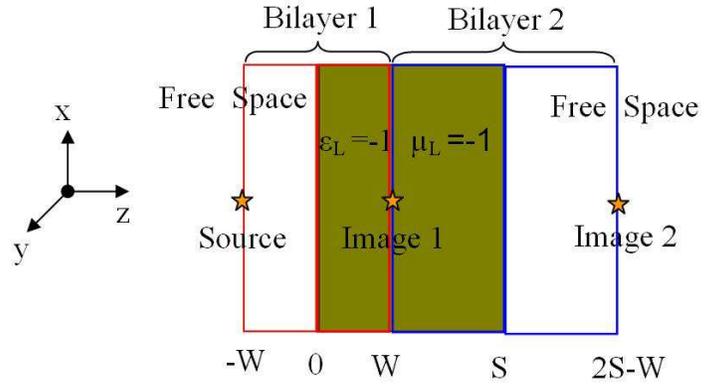}
\caption{Illustration of Pendry's perfect lens from the view of the
bilayer structure.}
\end{figure}
\newpage
\begin{figure}[htbp] \centering
\includegraphics[width=7cm]{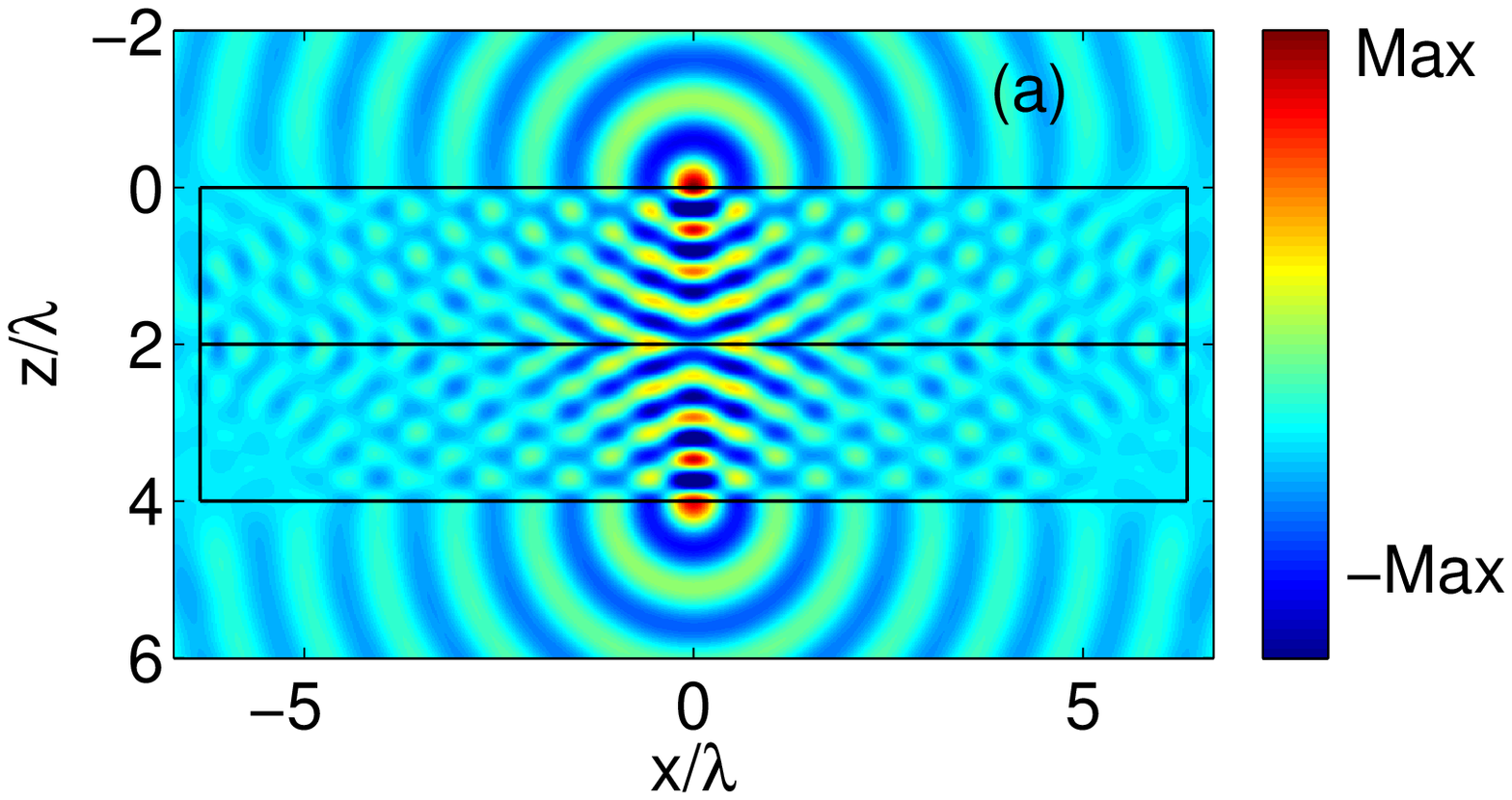}
\includegraphics[width=7cm]{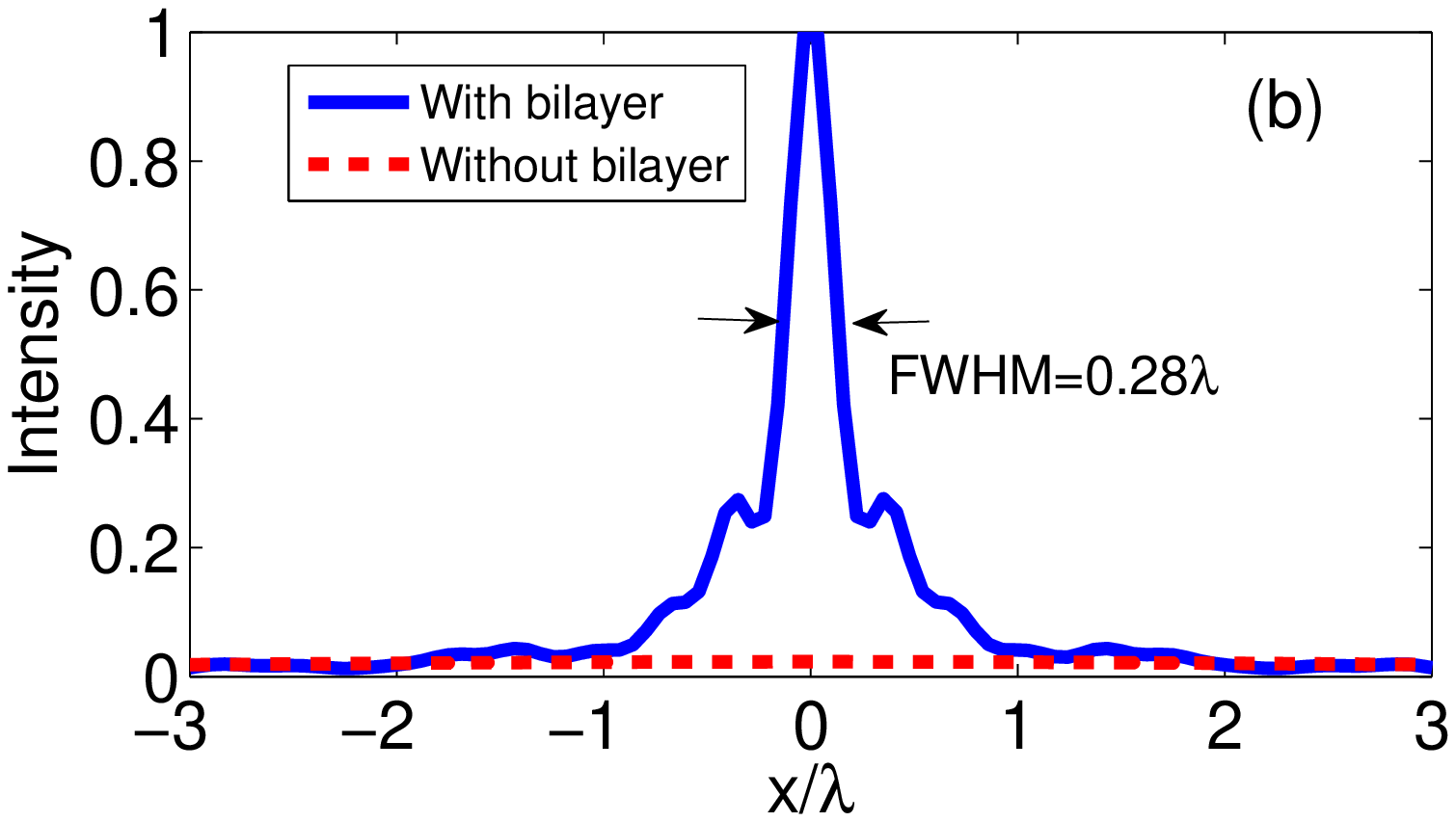}
\caption{(a) Electric field distribution for a line current source
$J_s=A\delta(z+0.01\lambda)\delta(x)$ interacting with a bilayer
lens located in $z\in[0, 4\lambda]$; (b) Field intensities at the
exit boundary $z=4\lambda$ for the cases with the bilayer and
without bilayer. The bilayer lens has $\epsilon_1=4$, $\mu_1=1$,
$\epsilon_2=-4+0.008i$ and $\mu_2=-1+0.008i$ and
$L_1=L_2=2\lambda$.}
\end{figure}
\newpage
\begin{figure}[htbp] \centering
\includegraphics[width=7cm]{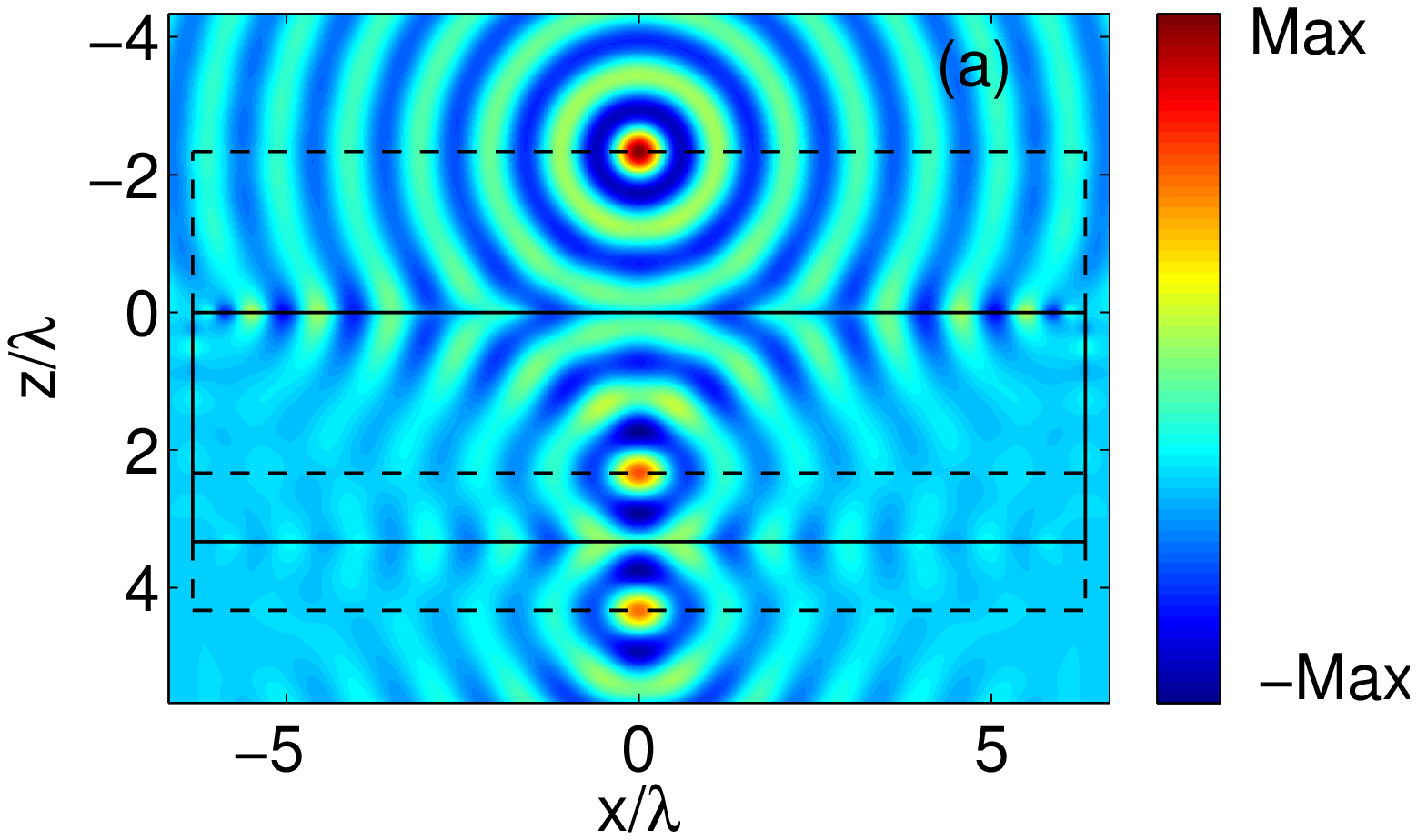}
\includegraphics[width=7cm]{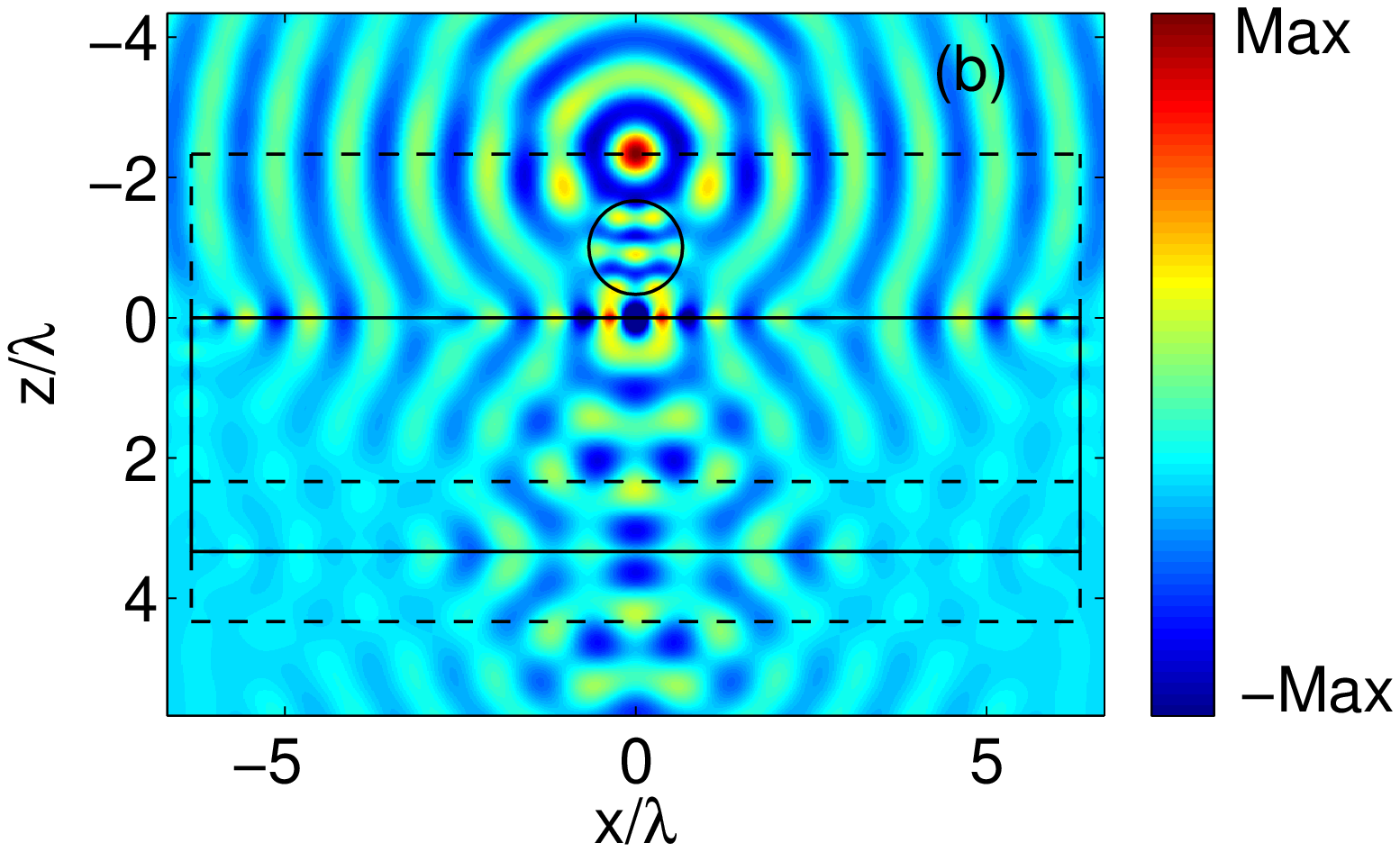}
\includegraphics[width=7cm]{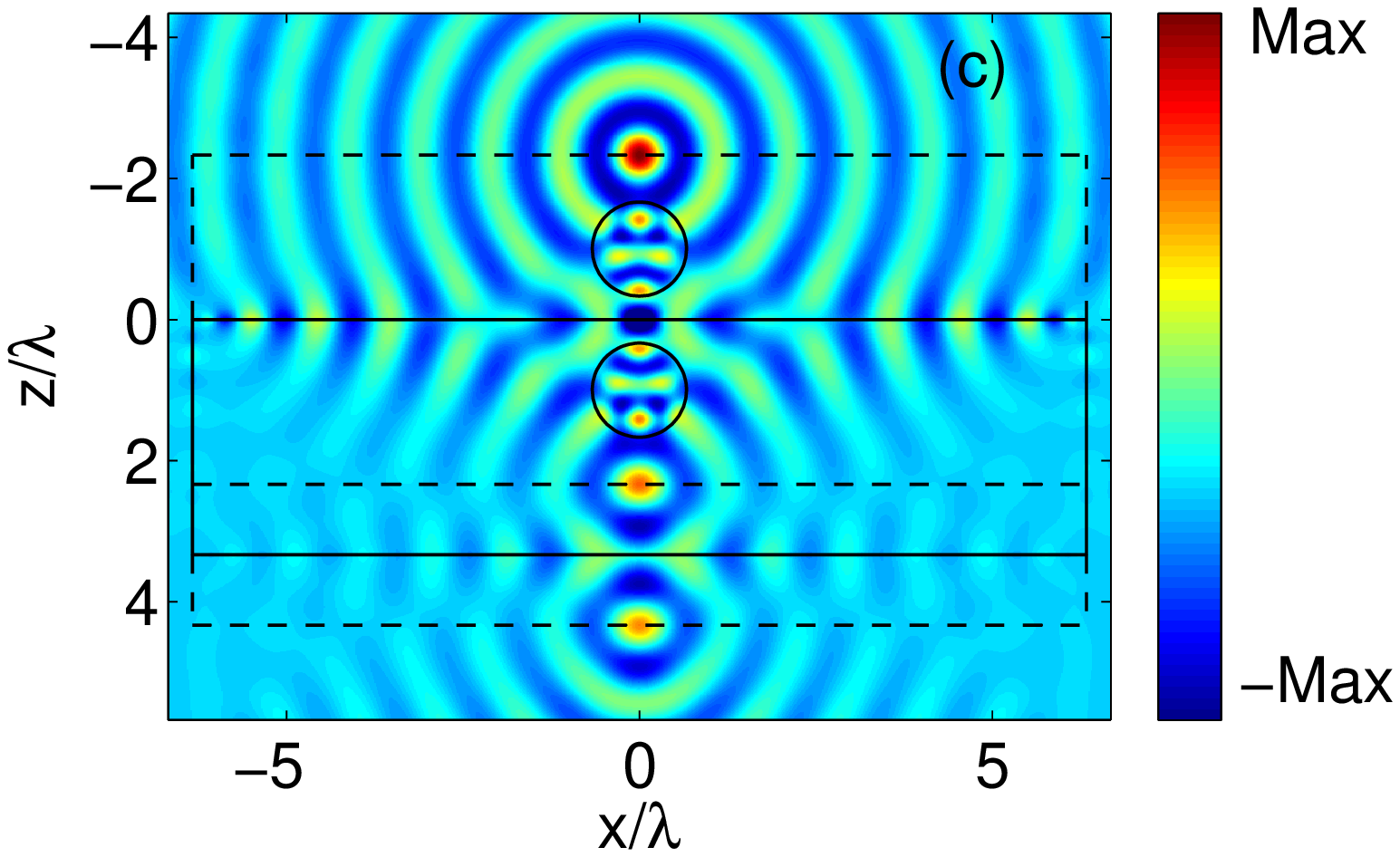}
\includegraphics[width=7cm]{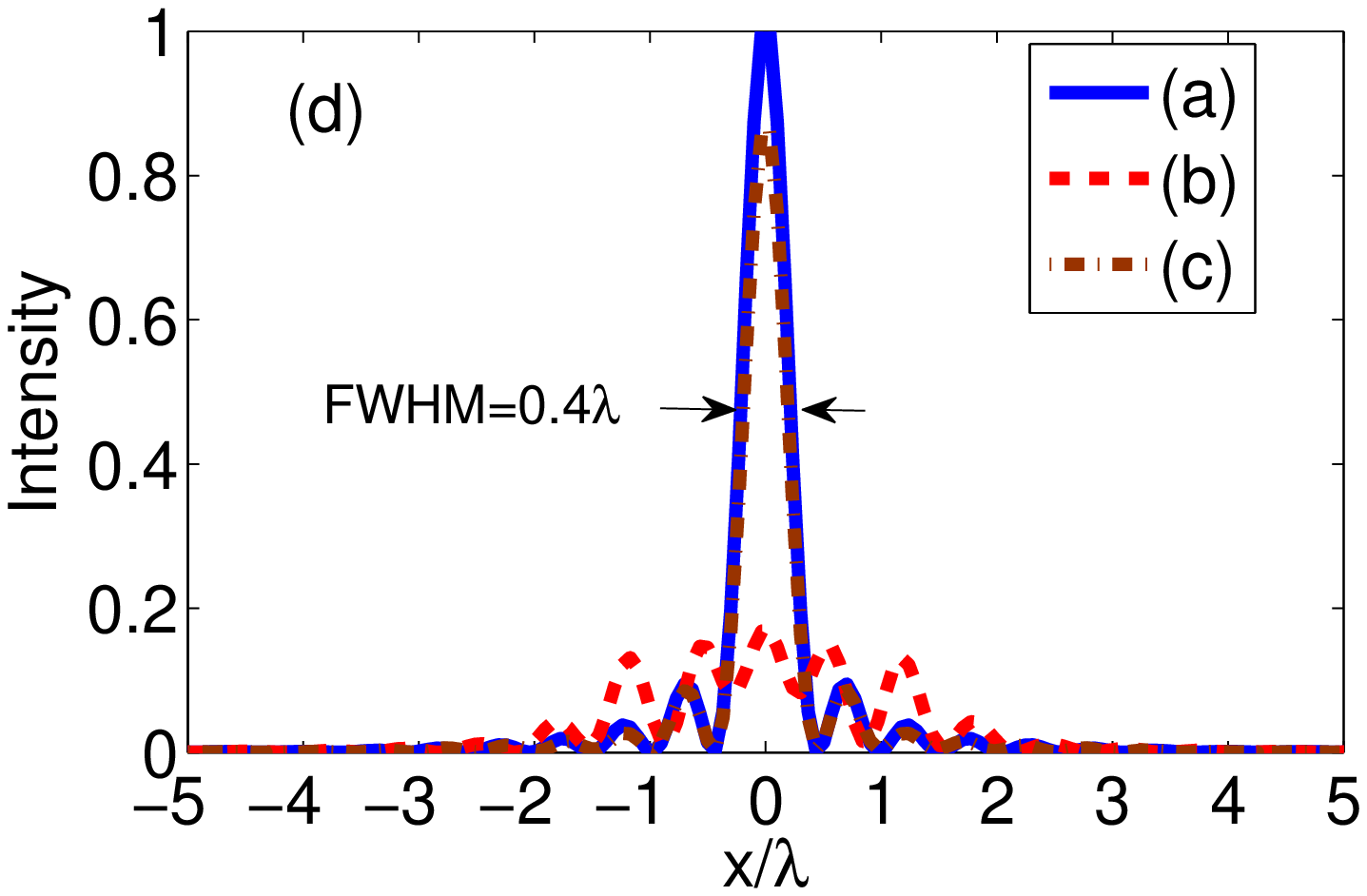}
\caption{(a) Electric field distribution for a line current source
$J_s=A\delta(z+7/3\lambda)\delta(x)$ interacting with a Pendry's
lens $\epsilon_L=-1+0.008i$ and $\mu_l=-1+0.008i$; (b) electric
field distribution when a dielectric cylinder is put in the front of
the lens; (c) electric field distribution when the dielectric
cylinder together with its compensated cylinder are put outside and
inside the lens, respectively. (d) Field intensity at the image
plane $z=13/3\lambda$ for (a), (b) and (c). The dielectric cylinder
has a radius of $2/3\lambda$ with the center at $z=-\lambda$ and
$x=0$, and a refractive index of $2$.}
\end{figure}
\newpage
\begin{figure}[htbp] \centering
\includegraphics[width=7cm]{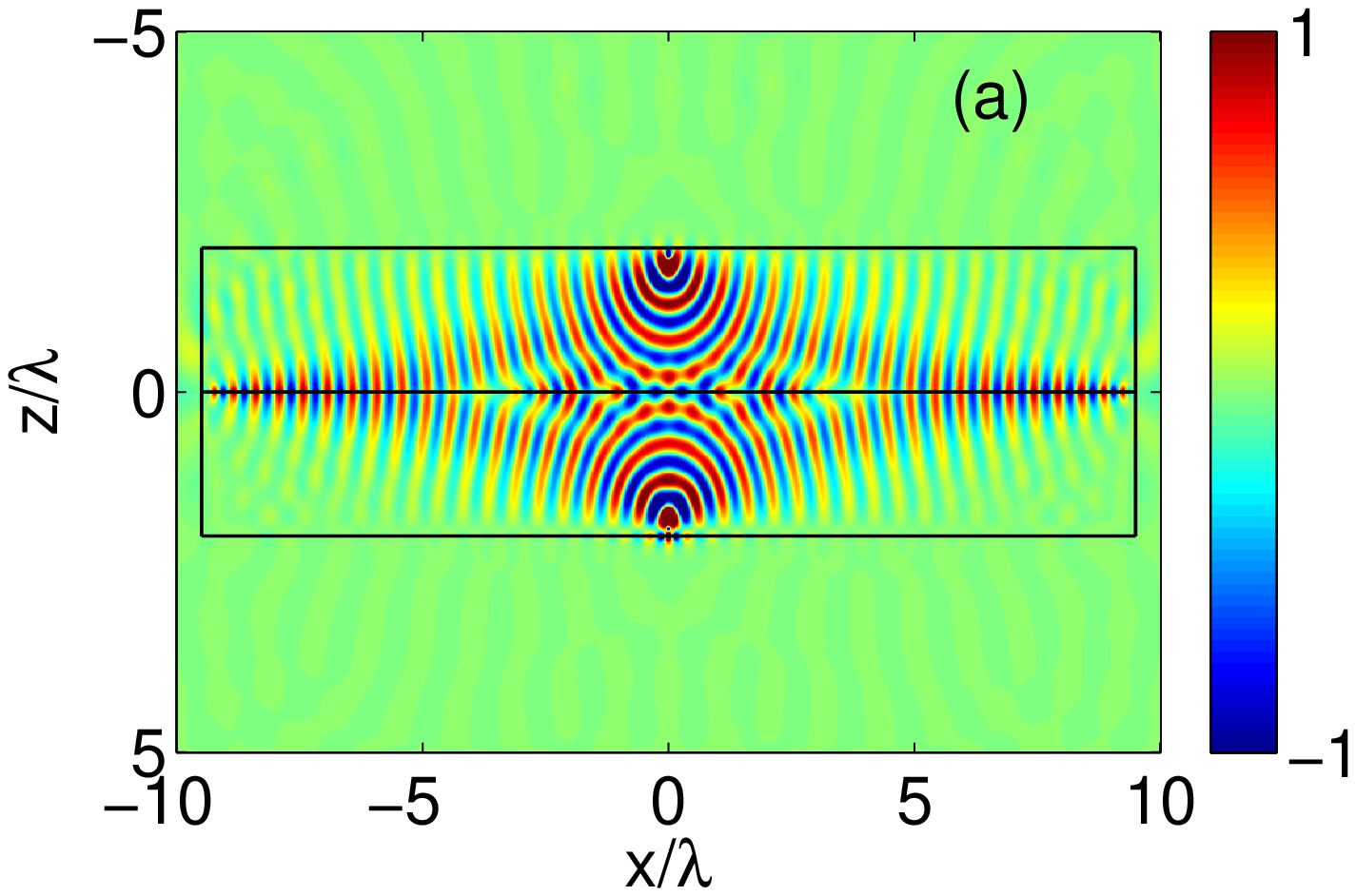}
\includegraphics[width=7cm]{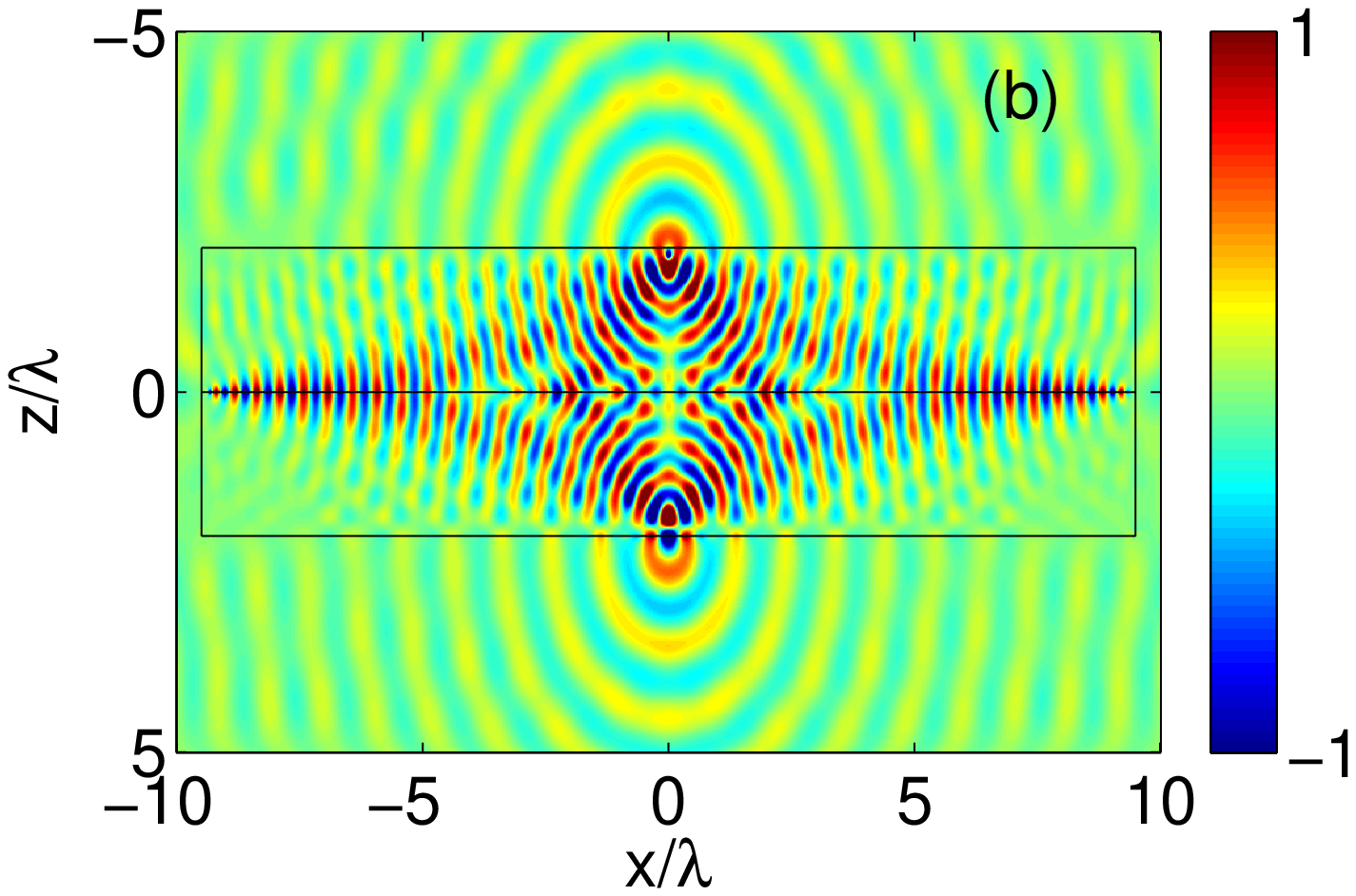}
\caption{Electric field distribution for (a) two current sources
$\overline J_1=A\delta(x)\delta(z+1.9\lambda)\hat y$ and $\overline
J_2=-A\delta(x)\delta(z-1.9\lambda)\hat y$ embedded in the bilayer ;
(b) only $\overline J_1$ is embedded in the bilayer. The bilayer
lens has the same material parameters as in Fig. 3.}
\end{figure}
\newpage
\begin{figure}[htbp] \centering
\includegraphics[width=7cm]{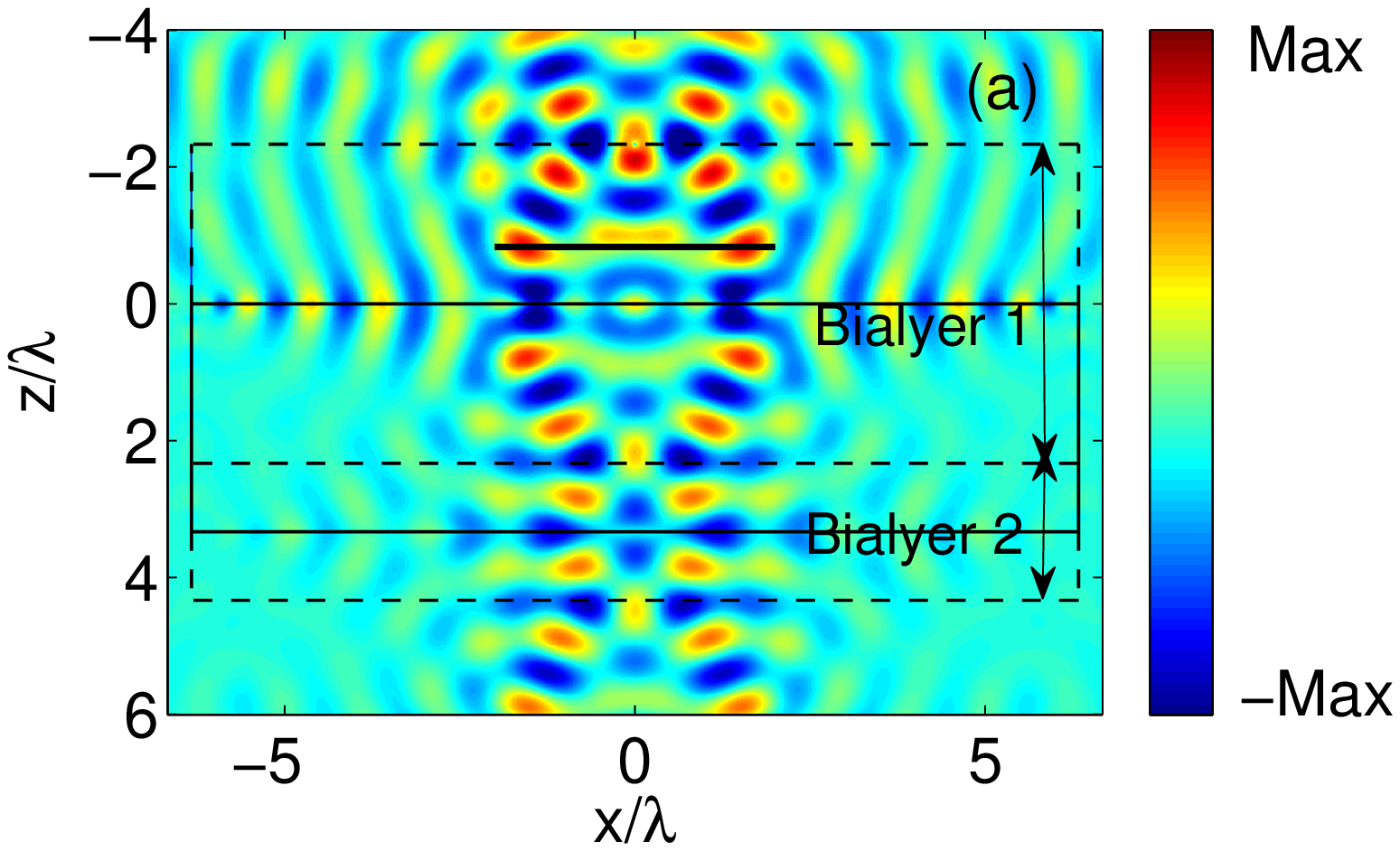}
\includegraphics[width=7cm]{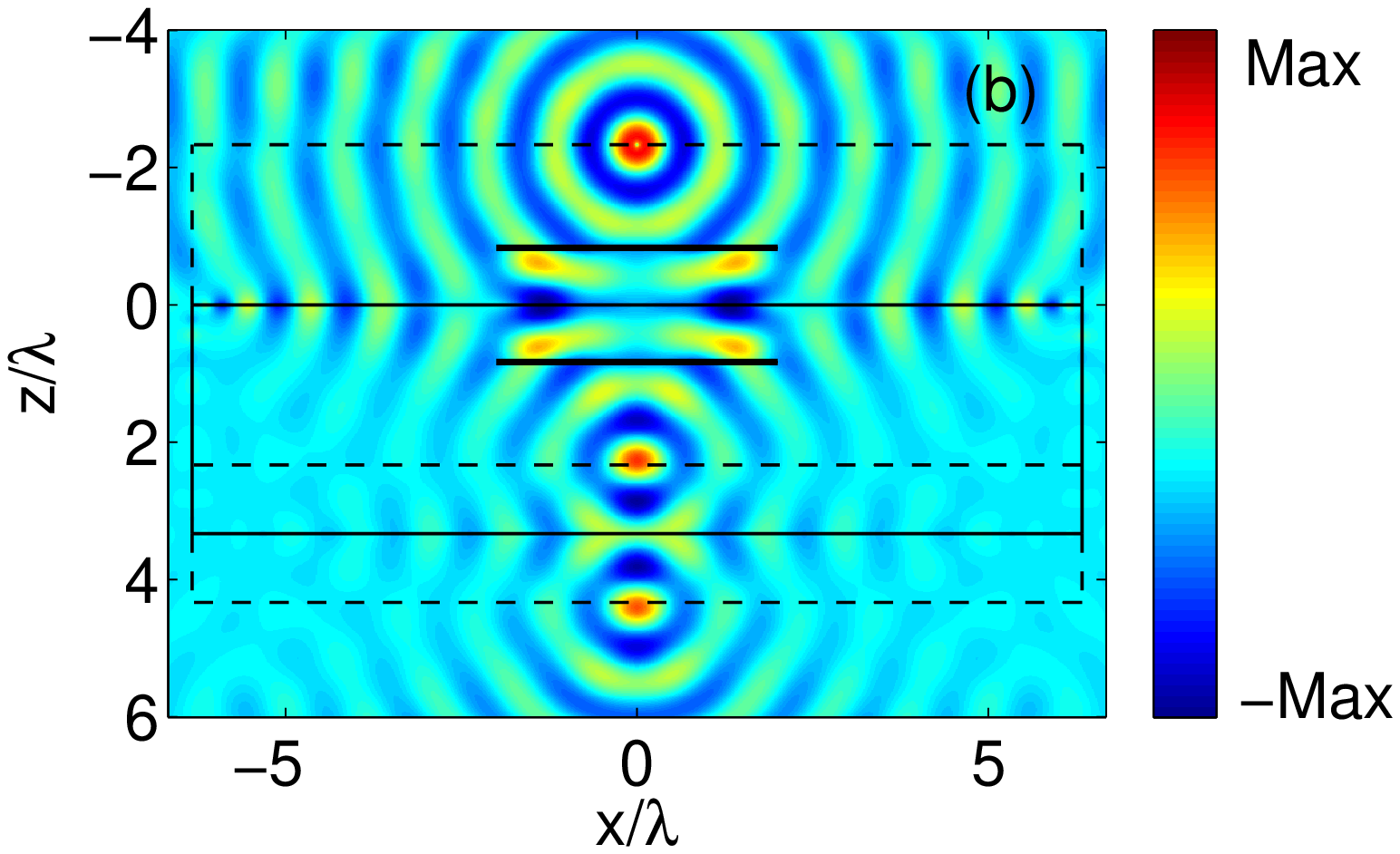}
\includegraphics[width=7cm]{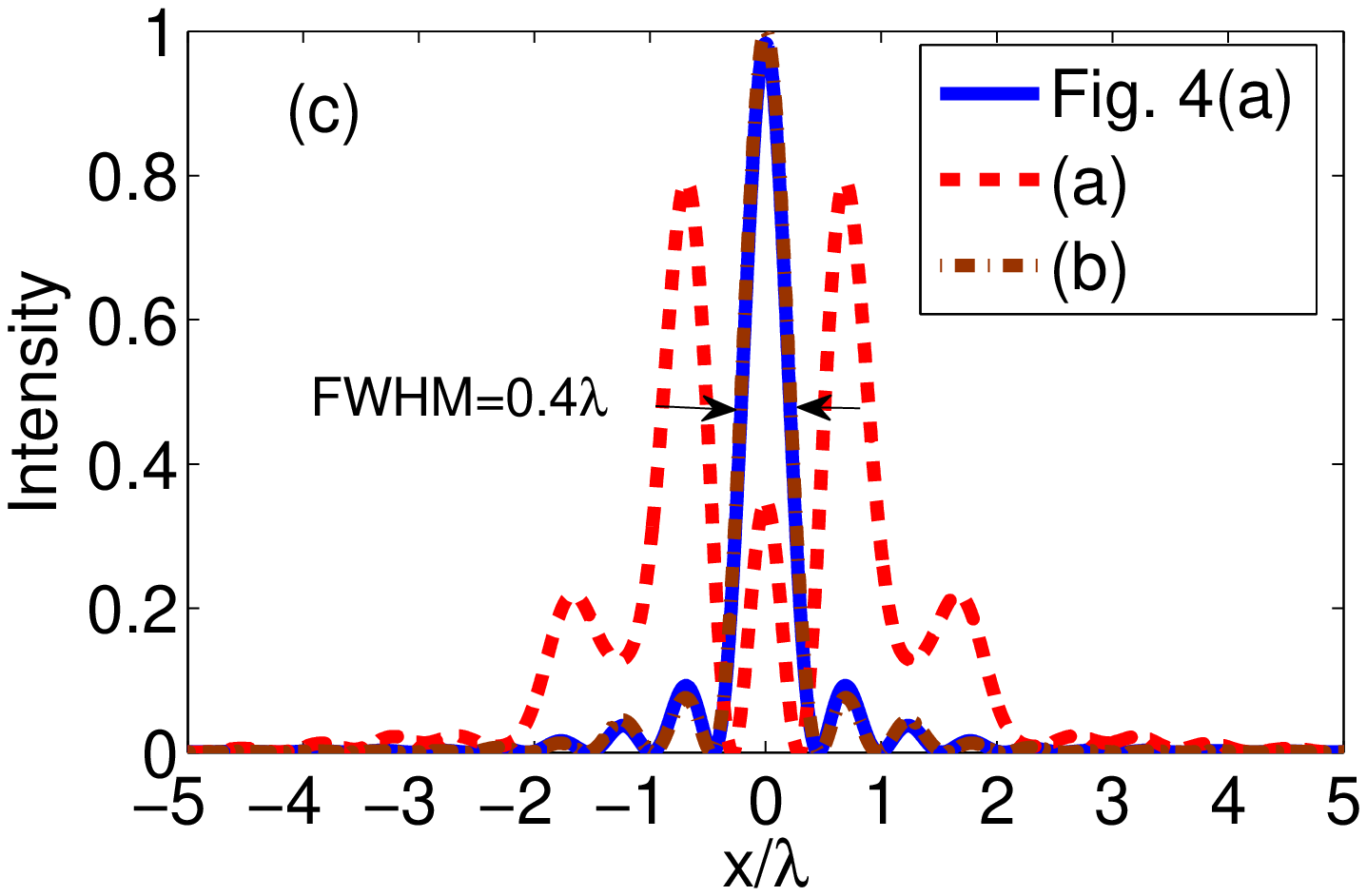}
\caption{Electric field distribution when (a) a current sheet
$J_s=4A\delta(x+z/6\lambda)\;x\in[-2\lambda,2\lambda]$ is put in the
front of the lens of Fig. 4(a); (b) the compensated current sheet is
put inside the lens. (c) Field intensity at the the image plane
$z=13/3\lambda$ for (a), (b), and also Fig. 4(a).}
\end{figure}

\newpage
\section*{References}
\begin{thebibliography}{}
\bibitem{P1} J. B. Pendry, D. Schurig, and D. R. Smith, Science
{\bf312}, 1780 (2006).
\bibitem{Ulf1} U. Leonhardt, and T. G. Philbin 2008 http:
//www.arXiv:0805.4778v2 [physics.optics].
\bibitem{Ulf2} U. Leonhardt, Science {\bf312}, 1777
(2006).
\bibitem{Sch}D. Schurig, J. J. Mock, B. J. Justice, S. A. Cummer, J. B.
Pendry, A. F. Starr, and D. R. Smith, Science {\bf314}, 977 (2006).
\bibitem{Cai} W. Cai, U. K. Chettiar, A. V. Kildishev, V. M.
Shalaev, Nat. Photonics {\bf 1}, 224-227 (2007).
\bibitem{Cum} S. A. Cummer, B. I. Popa, D. Schurig, D. R. Smith,
and J. B. Pendry, Phys. Rev. E. {\bf74}, 036621 (2006).
\bibitem{Ruan} Z. C. Ruan, M. Yan, C. W. Neff, and M. Qiu, Phys. Rev. Lett. {\bf99}, 113903 (2007).
\bibitem{Chen1} H. S. Chen, B. I. Wu, B. L. Zhang, and
J. A. Kong, Phys. Rev. Lett. {\bf99}, 063903(2007).
\bibitem{Ulf3} U. Leonhardt, New J. Phys. {\bf 8}, 247 (2006).
\bibitem{Yan} M. Yan, Z. C. Ruan, and M. Qiu, Phy Rev. Lett. {\bf
99}, 233901(2007).
\bibitem{Rahm} M. Rahm, S. A. Cummer, D. Schurig, J. B. Pendry, and
D. R. Smith, Phys. Rev. Lett. {\bf 100}, 063903 (2008).
\bibitem{Chen2} H. Y. Chen and C. T. Chan, Appl. Phys. Lett.
{\bf90}, 241105 (2007).
\bibitem{Gf} A. Greenleaf, Y. Kurylev, M. Lassas, and G.
Uhlmann, Phys. Rev. Lett. {\bf 99}, 183901 (2007).
\bibitem{P2} J. B. Pendry, Phys. Rev. Lett. {\bf85}, 3966-3969 (2000).
\bibitem{YM2} M. Yan, W. Yan, and M. Qiu, Phys. Rev. B {\bf 78}, 125113
(2008).
\bibitem{P3} J.B. Pendry, and S.A. Ramakrishna, J. Phys. Condensed Matter. {\bf 14}, 6345 (2003).
\bibitem{Smith}D.R. Smith, and S. Schurig, Phys. Rev. Lett. {\bf 90}, 077405 (2003).
\bibitem{Sch2}D Schurig and D R Smith, New Journal of Physics {\bf 7}, 162 (2005).
\end{thebibliography}
\end{document}